\documentclass[aip, apl, reprint, longbibliography]{revtex4-1}
\usepackage{hyperref}
\usepackage{graphicx}
\graphicspath{{figures/}}
\usepackage{amsmath}
\usepackage{amssymb}

\begin{document}

\title{Polarization-resolving graphene-based mid-infrared detector}

\author{Valentin Semkin}
\author{Dmitry Mylnikov}
\author{Elena Titova}
\author{Sergey Zhukov}
\author{Dmitry Svintsov}
\email[]{svintcov.da@mipt.ru}

\affiliation{Center for Photonics and 2D Materials, Moscow Institute of Physics and Technology, Dolgoprudny 141700, Russia}

\date{\today}

\begin{abstract}
The ability to resolve the polarization of light with on-chip devices represents an urgent problem in optoelectronics. The detectors with polarization resolution demonstrated so far mostly require multiple oriented detectors or movable external polarizers. Here, we experimentally demonstrate the feasibility to resolve the polarization of mid-infrared light with a single chemical-vapor-deposited graphene-channel device with dissimilar metal contacts. This possibility stems from an unusual dependence of photoresponse at graphene-metal junctions on gate voltage and polarization angle. Namely, there exist certain gate voltages providing the polarization-insensitive signal; operation at these voltages can be used for power calibration of the detector. At other gate voltages, the detector features very strong polarization sensitivity, with the ratio of signals for two orthogonal polarizations reaching $\sim 10$. Operation at these voltages can provide information about polarization angles, after the power calibration. We show that such unusual gate- and polarization-dependence of photosignal can appear upon competition of isotropic and anisotropic photovoltage generation pathways and discuss the possible physical candidates.
\end{abstract}

\maketitle

Light polarization brings important information about objects and materials not contained in color and intensity~\cite{Polarisation-imaging-1,Polarisation-imaging-2}. In modern optical communication systems, both classical and quantum, polarization state of light represents a new degree of freedom that can be used for information encoding~\cite{Polarisation-communications-1,Polarisation-communications-2,ran_integrated_2021}. Optoelectronic read-out of polarization state therefore represents an urgent technological problem. The absorbance of light-detecting elements generally depends on polarization via material~\cite{P-sensitive-material-1,P-sensitive-material-2,P-sensitive-material-3,tong_stable_2020} or structure~\cite{Grating-plasmon-detector,Grating-based-QWIP,wei_mid-infrared_2021,wei_zero-bias_2020} anisotropy. At the same time, development of polarization-{\it resolving} detectors is more challenging compared to polarization-{\it sensitive} ones, as it is necessary to disentangle the information about electromagnetic intensity and polarization from the detector signal.

A conventional approach for polarization resolution lies in a combination of detector and polarizer, either mechanically- or electrically-controllable~\cite{Micro-polarizer-array}. Both approaches are relatively slow, and some light power is inevitably lost due to polarizer. It is possible to get rid of polarizers by using sets of detectors with strongly polarization-sensitive absorption, oriented at various angles. Grating-based~\cite{Grating-plasmon-detector}, orthogonal-antenna-based~\cite{Mohammadi}, and nanowire-based~\cite{Nanowire-detector-1,Nanowire-detector-2} detectors are most recognized solutions in this regard. By normalizing the signals of all detectors in a set to the maximum, one gets rid of power dependence of signal. Comparison or normalized signals within a set is then used to derive the polarization angle. At least three detectors~\cite{Mohammadi} are required to uniquely determine the polarization angle.

To minimize the number of detectors in a polarization-resolving pixel and simplify the circuitry, it was recently proposed to use an electrical 'control signal' which strongly changes the angle-dependence of photoresponse~\cite{Polarization-resolved-bias-selectable}. In Ref.~\onlinecite{Polarization-resolved-bias-selectable}, this dependence was essentially different for forward- and backward-biased photodiode based on black phosphorous/molybdenum disulphide 2d heterostructure (bias-selectable detector). After adding a proper control signal, just two orthogonal detectors are sufficient for design of polarization-resolving pixel. Of course, using bias current as a 'control' leads to large shot noise and increases the noise equivalent power.

\begin{figure*}
    \includegraphics{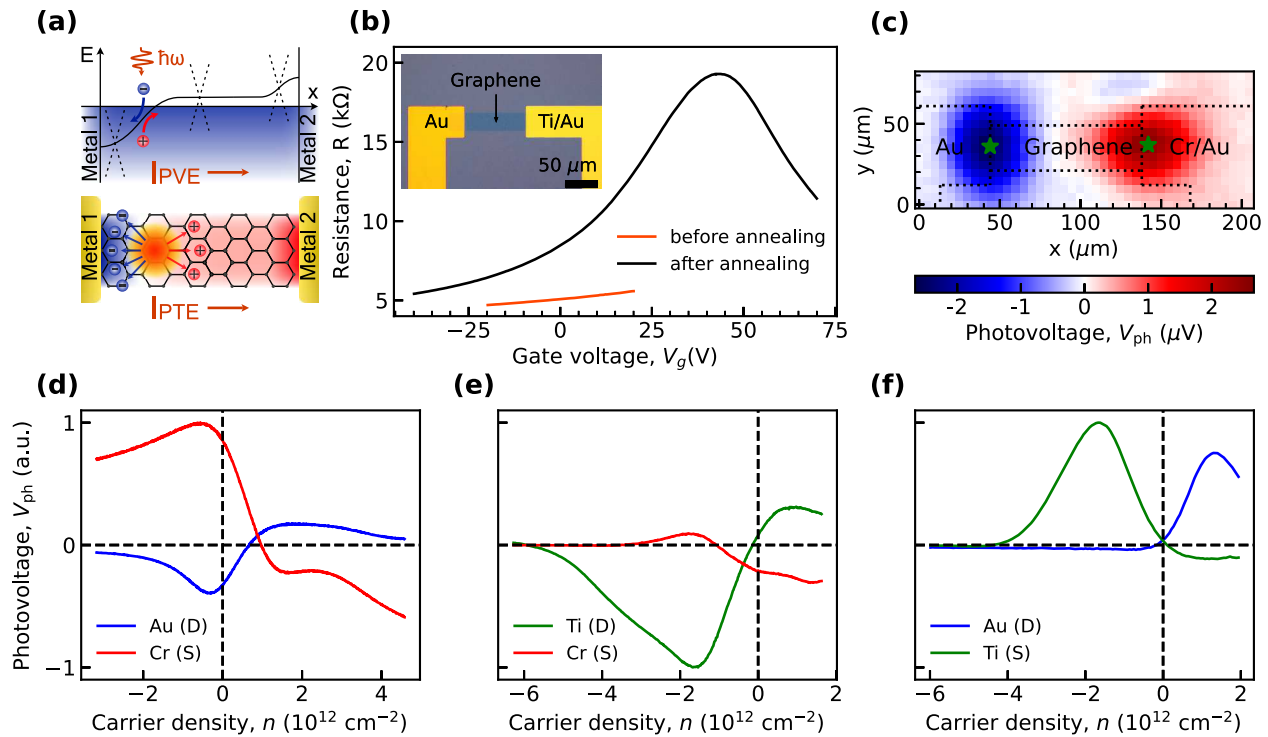}
    \caption{\label{fig1} (a) Schematic representation of the photovoltaic (PVE) and photo-thermoelectric (PTE) photovoltage generation mechanisms in graphene channel with dissimilar metal contacts. (b) Gate dependencies of the resistance for the graphene channel with Ti and Au contacts before and after current annealing. Inset: photo of the Ti-Au device. (c) Map of the photovoltage for the graphene device with Cr and Au contacts. The dotted lines represent the device edges and the stars represent the points at which gate dependencies of photovoltage were measured. (d)-(f) Gate dependencies of the photovoltage (at its maximum points) for Au-Cr (d), Ti-Cr (e) and Ti-Au (f) devices. 'S' and 'D' in the legend stand for source and drain; during measurements, the 'S' contact is grounded, the photovoltage is read out at the 'D' contact.}
\end{figure*}

Here, we propose and experimentally substantiate the operation of polarization-resolving detector where gate voltage is used as a control signal. Our detector is based on chemical vapor deposited graphene with two dissimilar metal contacts. Generation of photovoltage occurs at graphene-metal junctions via thermoelectric~\cite{Song,Tielrooij}, photovoltaic~\cite{Echtermeyer,Xia,Mueller}, plasmonic~\cite{Silkin,Bandurin_Svintsov}, or other unconventional~\cite{Ganichev_edge_BLG,Ganichev_edge_SLG} mechanisms. Independent of precise detection physics, dissimilarity of graphene-metal junctions leads to non-compensating partial photovoltages upon uniform illumination~\cite{Cai_14} [see Fig.~\ref{fig1}(a)]. 
The response of an individual graphene-metal junction was recently shown to be polarization-sensitive in the visible~\cite{Echtermeyer} and near-infrared~\cite{Tielrooij} ranges. This sensitivity was explained by the lightning-rod effect at metal junction for ${\bf E}$-field of the wave normal to the junction. For  ${\bf E}$-field parallel to the junction, the photovoltage was reduced compared to the orthogonal case, though having the same sign.

In this paper, we show that gate- and polarization-dependent response of graphene-metal junctions at the mid-infrared is more peculiar, and this enables polarization-resolving action. Namely, we show the presence of carrier density $n^*$ (set by the gate) at which the detector response is insensitive to the light polarization. This working point can be used for power calibration of the device. At other carrier densities ($n>n^*$ or $n<n^*$, depending on metal), the photoresponse has strong polarization sensitivity. The photovoltage ratio for two orthogonal polarizations reaching $\sim 10$. Operation at these densities, after power calibration, can be used for determination of the polarization angle $\theta$.

We argue that the polarization-insensitivity point $n^*$ appears due to the presence of two detection pathways, isotropic and aninsotropic, with opposite signs of photovoltage. Remarkably, there exists a range of densities, where $90^\circ$-rotation of polarization leads to sign flip of photovoltage. Both phenomena can be explained by the presence of two competing photovoltage generation mechanisms.


Our devices are based on commercial CVD graphene grown on Cu foil. Graphene was wet-transferred on oxidized (300 nm of SiO$_2$) Si substrate acting as a back gate. After that, graphene was patterned in oxygen plasma and contacts were fabricated using magnetron sputtering. The optical image of one of the devices is shown in the inset of Fig.~\ref{fig1}(b). The channel width in all our devices is 30 $\mu$m and the length is 100 $\mu$m.  It enables to record well-resolved photovoltage maps at a wavelength $\lambda = 8.6$ $\mu$m (generated by a quantum cascade laser with power $P_{\rm las}\approx 20$ mW). As the FWHM of the focused laser beam being $\sim 30$ $\mu$m is less than device length, we were able to illuminate and study graphene-metal junctions independently of each other.

Because of the large initial doping of non-encapsulated CVD graphene, the first step in our measurements is annealing. We have used current annealing of devices in vacuum to get rid of contaminants and to move the charge neutrality point (CNP) closer to the zero gate voltage. The gate dependencies of the resistance for Ti-Au device before and after annealing are shown in the Fig.~\ref{fig1}(b). Transport characterizations were carried out with about 10 mV DC bias on two-channels source-meter.

A simple graphene channel with identical metal contacts does not generate photovoltage at zero bias and uniform illumination. The reason lies in opposite signs of photovoltage at source-channel and channel-drain junctions. To circumvent this problem, we introduce lateral asymmetry to the device by fabricating contacts of dissimilar metals~\cite{Cai_14,Fedorov_asymmetric}. Ideally, the source and drain metals should dope graphene with carriers of opposite polarity, thus the photovoltages at two junctions should sum up.

The type of metal-induced doping is mainly determined by the work functions of metals~\cite{Metal-graphene-doping-1}. However, it is also affected by the sputtering conditions, substrate, and possible contaminants trapped upon wet transfer~\cite{Metal-graphene-doping-2}. To make a practical selection of optimal contact pair, choose metals and maximize the detector efficiency, we have fabricated a series of devices with various metals: Au, Cr and Ti. The optimal pair was selected based on the photovoltage measurements.

The first step in optical measurements was to obtain the spatial maps of the arising photovoltage $V_{\rm ph}$ [Fig.~\ref{fig1}(c)]. They were used to align the experimental setup, as well as to determine the positions of the graphene-metal junctions. During all measurements, the sample was held in an evacuated cryostat at room temperature. Generated photovoltage was measured by a lock-in amplifier. 

After the graphene-metal junctions were identified optically, the dependencies of the photovoltage on the charge carrier density $n=n_e-n_h$ in the channel were obtained by tuning the gate voltage [Fig.~\ref{fig1}(d)-(f)], $n_e$ is the electron density, $n_h$ is the hole density. We observe that signals generated by Au and Cr junctions, Ti and Cr junctions tend to compensate each other. The device with Ti and Au junctions turns out to be optimal. Although there is no photovoltage enhancement at sequential contacts, but more importantly, there is no tangible photovoltage compensation either. Indeed, graphene-Au junction generates large photovoltage at electron doping of the channel, while the graphene-Ti junction photovoltage is negligible here. The situation at hole doping is just the opposite: graphene-Ti junction generates large positive photovoltage, while graphene-Au junction is almost light-insensitive. The above reasoning shows that Ti-Au contact pair is most efficient and further measurements were performed for it.

\begin{figure}
    \includegraphics{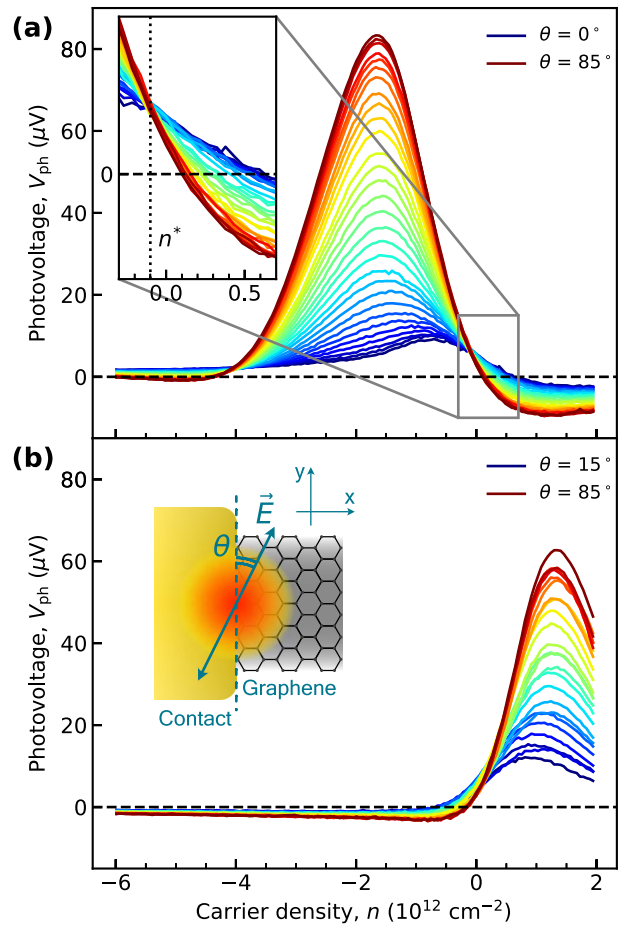}
    \caption{\label{fig2} Dependencies of photovoltage on carrier density and polarization angle at graphene-Ti (a) and graphene-Au (b) junctions. The polarization angle $\theta$ is counted from the direction that parallel to the graphene-metal junction (bottom inset), and varies with $3^\circ$ step. Inset in (a) magnifies the vicinity of carrier density $n^*$ where the photoresponse becomes polarization-independent.}
\end{figure}

Having selected the contact pair, we proceed to the investigation of the polarization sensitivity of junctions. For this purpose, we place a quarter-wavelength plate and polarizer between the focusing lens and the sample. After the quarter-wavelength plate we get circular-polarized light, since our laser generates linear-polarized radiation and the optical axes of the plate are set at $+\pi/4$ and $-\pi/4$ relative to it. Then we can rotate the incident radiation polarization using the polarizer, but power remains constant. Illuminating each junction independently of the second one again and rotating the polarization, a series of dependencies of $V_{\rm ph}$ on polarization angle $\theta$ and carrier density $n$ were obtained [Fig.~\ref{fig2}].

The first remarkable property of the photoresponse curves is their asymmetry with respect to the carrier density. The photovoltage is large only on one side from the CNP, and these sides are different for Ti and Au. We attribute the enhancement of photoresponse to the formation of $p-n$ junctions at graphene-metal interfaces~\cite{Tielrooij}. The polarity of near-contact part of the junction is controlled by the type of metal, while the polarity of the opposite part is controlled by the bottom gate. Based on this argument, we can conclude that in our device Ti dopes graphene with electrons and Au dopes graphene with holes. The latter fact is in agreement with work function arguments~\cite{Metal-graphene-doping-1}.

The second remarkable property of the photoresponse is its strong polarization sensitivity of both graphene-Ti [Fig.~\ref{fig2}(a)] and graphene-Au [Fig.~\ref{fig2}(b)] junctions. The photovoltage increases by an order of magnitude when we rotate the ${\bf E}$-field of radiation from parallel to junction ($\theta = 0$) to the perpendicular ($\theta = 90^\circ$). One of the origins for strong polarization sensitivity is the lightning-rod effect, the enhancement of local electric field by keen metal objects~\cite{Sommerfeld,Nikulin}. This is indeed our case, as metal contacts to graphene have thickness $t\sim 70$ nm well below the radiation wavelength. The diffraction of electromagnetic waves with ${\bf E}$-field perpendicular to the edge results in singular enhancement of local field near the edge, $E_{\rm loc, x}\propto x^{-1/2}$, where $x$ is the distance to the edge~\cite{Nikulin}. No such enhancement should occur for the component of ${\bf E}$-field parallel to the edge, the respective local field $E_{\rm loc, y}$ remains order of incident field $E_0$ in immediate vicinity of the junction. The solution of diffraction problem for two orthogonal polarizations is shown in the Fig.~\ref{fig3} and detailed in Supplementary Section~I.
\begin{figure}
    \includegraphics{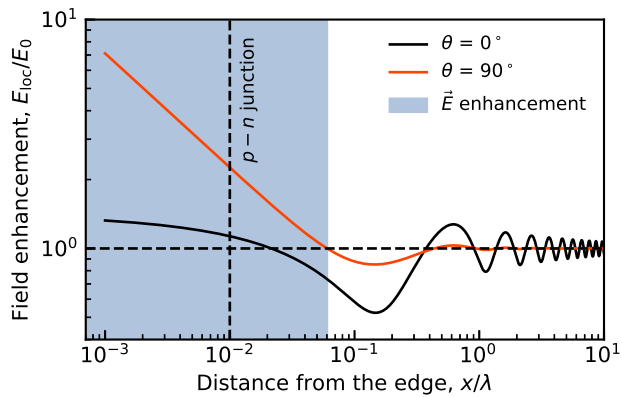}
    \caption{\label{fig3} Calculated coordinate dependence of local electric field at graphene-metal interface for two orthogonal polarizations of incident light: $\bf E$-field parallel to the contact (black) and $\bf E$-field normal to the contact (orange). In the blue shaded area, the amplitude of local field exceeds the incident field due to the lightning-rod effect. Approximately at $x\sim 100$ nm, a $p-n$ junction due to metal-induced doping is expected.}
\end{figure}
At distance $x\sim 100$ nm, which is approximately the length of metal-induced $p-n$ junction (Supplementary Section II), the local field for $\theta = 0^\circ$ is roughly three times larger than that for $\theta = 90^\circ$. This is, in principle, sufficient to explain the order-of-magnitude ratio of local absorbed powers, and respective photovoltages.

It is important that the polarization contrast in our case is stronger than in Refs.~\onlinecite{Echtermeyer, Tielrooij} that dealt with visible/near-infrared light. The difference can be also explained by the lightning-rod effect. As evident from scaling arguments (and from exact solution of the diffraction problem), the spatial extent of edge-enhanced field is proportional to $\lambda$. Actually, the field enhancement ($E_{\rm loc}/E_0>1$) occurs at $x\lesssim 0.06\lambda \approx 500$ nm. The enhancement of photovoltage depends on whether the light-sensing $p-n$ junction overlaps with region of enhanced field or not. For mid-infrared light, the junction length $l_{p-n}\sim 100$ nm indeed fits into the region of edge-enhanced field. For visible light, this is already not the case, and polarization contrast of photovoltage is lower. Passing to numerical comparison, Ref.~\onlinecite{Echtermeyer} reported subsidence of the photovoltage at the parallel to contacts polarization by 20\%, 25\%, and 50\% relative to the perpendicular polarization for 633 nm, 785 nm and 1550nm wavelengths respectively. In our work, we observe a subsidence in the photovoltage by 95\% for excitation at 8.6 $\mu$m at the carrier density that provides the maximum responsivity. Similarly, Ref.~\onlinecite{Tielrooij} shows responsivity swing of 40\% relative to the central signal level at 630 nm, in our case it is about 100\%. Thus, the graphene-metal junction indeed demonstrates an increase in polarization contrast with an increase in the wavelength.

A more detailed inspection of photovoltage data reveals several features that cannot be explained by electrodynamics only, and require polarization sensitivity of microscopic detection mechanisms themselves. These features would allow the detector to resolve polarization. The region of interest is zoomed for Ti in Fig.~\ref{fig2}(a), inset. There exists a range of carrier densities in which the photovoltage changes sign upon polarization rotation. Variations of local electric field can only scale the signal, but not change its sign. None of established detection mechanism has this property of sign flip upon polarization rotation: PTE is isotropic, as electronic heating is insensitive to local field direction; PVE is anisotropic, but its sign depends on the direction of built-in field.

In addition to a range of densities with sign flip of photovoltage upon polarization rotation, we mention the presence of the 'fixed points' $n^*$. More precisely, at these carrier densities the photovoltage does not depend on the incident radiation polarization. An important role of these points for polarization resolution will be explained below. The physics beyond polarization-insensitivity points, as well as the physics beyond sign flips, may lie in the presence of at least two detection mechanisms that have different signs of responsivity and different dependence on local polarization angle. In the limiting case, one of the mechanisms is fully isotropic, while the other reacts only to normal component of electric field $E_{\rm loc, x}$. Denoting the local field enhancement factors for $x$- and $y$-polarized light as $k_x$ and $k_y$, we can present the full photovoltage as:
\begin{multline}
\label{v(iso,aniso)}
	V_{\rm ph} = R_{\rm is}(E_{\rm loc, x }^2 + E_{\rm loc, y }^2) + R_{\rm an}E_{\rm loc, x }^2\\
	= E_0^2 \left\{ R_{ \rm  is}k_y^2 + \left[ R_{\rm is}(k_x^2- k_y^2) + R_{\rm an} k_x^2\right]\sin^2\theta \right\}\\
    \equiv V_{\rm is} + V_{\rm an}\sin^2\theta,
\end{multline}
where we have introduced the responsivities of isotropic and anisotropic mechanisms $R_{\rm is/an}$ being the proportionality coefficients between photovoltage and squared electric field.

The presence of two competing mechanisms, isotropic and anisotropic, can indeed explain the unusual polarization response. The signal at perpendicular polarization, in such model, will be $V_\bot \propto (R_{\rm is}+R_{\rm an})k_x^2$, while the signal at parallel polarization is $V_\parallel \propto R_{\rm is}k_y^2$. Naturally, if an anisotropic mechanism has opposite sign to the isotropic one, and is stronger in magnitude, $V_\bot \propto R_{\rm is}+R_{\rm an}$ can be opposite to $V_\parallel \propto R_{\rm is}$ in some range of carrier densities. Moreover, in the proposed model, the response should become polarization-insensitive if $R_{\rm is}(k_x^2- k_y^2) + R_{\rm an} k_x^2 = 0$. Provided that enhancement for normal field is large, $k_x/k_y \gg 1$, the polarization-insensitivity point occurs whenever $R_{\rm is}(n^*) = - R_{\rm an}(n^*)$. To conclude, both sign flip of the signal with polarization rotation and polarization-insensitivity points can be explained within a model with two detection mechanisms with opposite sign.

\begin{figure}
    \includegraphics{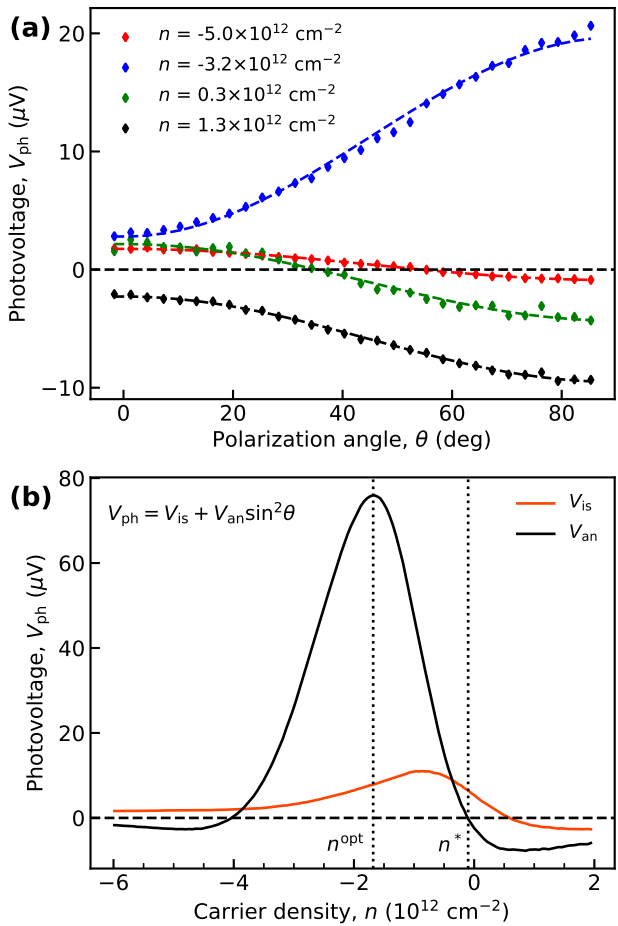}
    \caption{\label{fig4} (a) Polarization dependencies of photovoltage at four characteristic carrier densities for the graphene-Ti junction. Dashed lines represent the fit of photovoltage signal with $V_{\rm ph} = V_{\rm is} + V_{\rm an}\sin^2\theta$, meaning its splitting into isotropic and anisotropic components. (b) Dependencies of the isotropic $V_{\rm is}$ and anisotropic $V_{\rm an}$ contributions to the photovoltage on the carrier density for the graphene-Ti junction.}
\end{figure}

Our findings are further detailed in Fig.~\ref{fig4}(a), 
where we plot the dependencies of signal on polarization angle at various carrier densities for graphene-Ti junction. While $n=-3.2 \times 10^{12}$~cm$^{-2}$ (blue) and $n=1.3 \times 10^{12}$~cm$^{-2}$ (black) correspond to 'normal' polarization sensitivity, the curves for $n=-5.0 \times 10^{12}$~cm$^{-2}$ (red) and $n=0.3 \times 10^{12}$~cm$^{-2}$ (green) correspond to sign-flipping sensitivity. In the density range corresponding to 'normal' response, the polarization contrast of our detector is indeed large. This can be further substantiated by plotting the dependencies of 'isotropic' and 'anisotropic' photovoltage components $V_{\rm is}$ and $V_{\rm an}$ on carrier density, Fig.~\ref{fig4}(b). To the left from CNP, $V_{\rm an}$ exceeds $V_{\rm is}$ by an order of magnitude at its maximum.


The unique features of the polarization dependencies [Fig.~\ref{fig2}] and the linearity of the detector response to the radiation power $P$ allow us to determine $P$ and the polarization angle $\theta$ simultaneously. To do this, it's sufficient to measure the signal at two characteristic carriers densities. The first carrier density, $n^*$, provides the polarization-insensitive response, while the second, $n^{\rm opt}$, provides the strongest polarization sensitivity. The photovoltage $V(n^*)$ at the polarization-insensitive point $n^*$ is proportional to $P$, which allows us to determine $P~=~V(n^*)\frac{P_{\rm cal}}{V_{\rm cal}(n^*)}$, where $P_{\rm cal}$ is the power at which the calibration curves were measured, and $V_{\rm cal}(n^*)$ is the respective 'calibration photovoltage'. Normalization of the signal by $V(n^*)$ at other densities allows us to get rid of the power dependence and to determine $\theta$. For this purpose, it is rational to set the carrier density to $n^{\rm opt}$ that provides the maximum detector responsivity and its maximum variation with angle. Having measured the photovoltage $V(n^{\rm opt})$, one determines $\theta$ based on the known calibration fit parameters $V_{\rm is}$ and $V_{\rm an}$ from the equation:

\begin{equation}
    \dfrac{V(n^{\rm opt})}{V(n^*)}= \dfrac{ V_{\rm is}(n^{\rm opt}) + V_{\rm an}(n^{\rm opt})\sin^2\theta }{V_{\rm is}(n^*)}
\end{equation}

The presented technique is described for the case when $\theta$ is determined with respect to the selected direction, i.e. in the $\pi/2$-range. If the goal is to determine $\theta$ in the full $\pi$-range, then an array of two detectors can be used. They should be rotated relative to each other in such a way that the angle between the edges of their contacts is not a multiple of $\pi/2$, for example $\delta\phi~=\pi/4$. In this case, two detectors with a photovoltage dependence on polarization  of the forms $V_1=a_1+b_1\sin^2(\theta)$ and $V_2=a_2+b_2\sin^2(\theta-\pi/4)$ will parametrically set an ellipse $(V_1(\theta),V_2(\theta))$ with a $\pi$ period. Thus it becomes possible to resolve the radiation polarization in the full $\pi$-range.


In conclusion, we have experimentally substantiated a method to determine the light intensity and polarization from the gate-dependent photoresponse of graphene-metal junctions at mid-infrared wavelengths. A unique feature of angle- and density-dependent responsivity of such junctions lies in the existence of carrier densities at which the photoresponse becomes polarization-independent. At other carrier densities, the photoresponse may exhibit very strong polarization sensitivity (reaching $\sim$ 10 for two orthogonal polarizations). Addressing to these two carrier densities with global gate, one gets sufficient information to determine the light power and polarization angle within the $\pi/2$-range. Use of two mutually rotated detectors enables the unique specification of polarization angle within the $\pi$-range. The presence of polarization-insensitivity points can be explained by the presence of competing photovoltage generation pathways, one being isotropic and the other being anisotropic. Here, we did not attempt to fit the recorded responsivity curves with a combination of known detection mechanism. Yet we expect the isotropic part of response to originate from photo-thermoelectric effect, while the anisotropic photoresponse may stem from photovoltaics or plasmonic drag.

See the supplementary material for (I) details of modeling for polarization-dependent diffraction at graphene-metal contact (II) modeling of metal-induced doping in gated graphene.

This work was supported by the grant 21-72-00078 of the Russian Science Foundation (optical and electronic measurements) and by Ministry of Science and Higher Education of the Russian Federation (No.  FSMG-2021-0005) (device fabrication). The devices were fabricated using the equipment of the MIPT Center of Shared Research Facilities.

{\bf{Data availability statement.}} The data that support the findings of this study are available from the corresponding author upon reasonable request.

\bibliography{main}

\end{document}


\section{Polarization-dependent diffraction at metal contact}
In this section, we provide a sketch of solution for electromagnetic wave diffraction at a metal contact. The details of solution can be found in~\cite{Sommerfeld,Noble1958MethodsBO}, and its generalization to metal plane contacting graphene -- in Ref.~\onlinecite{Nikulin}.

The governing equation for light diffraction at a thin metal edge (occupying the half-plane $z=0$, $x<0$) is obtained from the following arguments. First, the total electric field at the metal edge is the sum of incident field ${\bf E}_0$ and the field created by the induced surface currents ${\bf j}(x)$. Second, the relation between induced field and the currents is written in integral form using the fundamental solution of the wave equation. Third, the total field (incident plus induced) at the metal contact should be identically zero. This leads us to the following equations (written separately for $x$- and $y$-polarizations):
\begin{gather}
E_{x}(x) = E_{0x}e^{i k_x x}-\frac{\pi}{k c}\left(k^{2}+\partial_{x}^{2}\right) \int\limits_{-\infty}^0 d x^{\prime} j_{x}(x') H_{0}(k|\Delta x|),\\
E_{y}(x) = E_{0y}e^{i k_x x}-\frac{\pi}{c} k \int\limits_{-\infty}^0 d x^{\prime} j_{y}(x') H_{0}(k|\Delta x|).    
\end{gather}
In the above equations, $k$ is the wave number of incident radiation, $H_0$ is the Hankel function. Equations are written down in Gaussian units.

Applying Fourier transform and assuming that the incident field exists only above the plane (this assumption will not affect the diffracted wave), we obtain the following transformed equations in the wave vector $q$-domain:

\begin{gather}
\label{F-transformed}
E_{x}(q)=-\frac{E_{0 x}}{i q - i k_x}-\frac{2 \pi}{k c} \sqrt{k^{2}-q^{2}} j_{x}(q), \\
E_{y}(q)=-\frac{E_{0  y}}{i q - i k_x}-\frac{2 \pi k}{c} j_{y}(q) \frac{1}{\sqrt{k^{2}-q^{2}}}
\end{gather}

The equations are subsequently solved with Wiener-Hopf method, which requires splitting of functions in Eqs.~(\ref{F-transformed}) into those analytic in the upper and lower half-planes. This results in:
\begin{gather}
\label{Transformed}
E_{x}(q)=\frac{i E_{0 x}}{\cos\theta - q/k}\left(1 - \frac{f(q/k)}{f(\cos \theta)} \right) \\
E_{y}(q)=\frac{E_{0 y}}{\cos\theta - q/k}\left(1 - \frac{f(\cos\theta)}{f(q/k)} \right),
\end{gather}
where $f(x)$ is the branch of the square root function $\sqrt{x+1}$ defined according to
\begin{equation}
    f(x) = \sqrt{x+1}\vartheta(x+1) + i \sqrt{-x - 1} \vartheta (-x-1),
\end{equation}
and $\vartheta$ is the Heaviside step function.

We apply the inverse Fourier transform to (\ref{Transformed}). After a change of variables, the final expressions for diffracted fields take the form:
\begin{equation}
\label{Ex}
    E_x(x) = E_{0x}\left\{e^{-i k x \cos (\theta )} + i e^{i k x}
     \left[\frac{e^{-i\pi/4}\sec \frac{\theta }{2}}{\sqrt{2\pi k x} } + i e^{- i k x (\cos \theta+1)} \left(1+\text{erf}\left(e^{i\pi/4} \sqrt{-2 k x} \cos
   \frac{\theta }{2}\right)\right)\right]\right\}
\end{equation}
\begin{equation}
\label{Ey}
    E_y(x) = E_{0y} e^{- i k x \cos\theta } \left\{i  \text{erf}\left(e^{i\pi/4}\sqrt{-2kx} \cos \frac{\theta}{2}\right)+ 1+i \right\}.
\end{equation}

The results in Fig.~3 of the main paper were obtained with Eqs.~(\ref{Ex}) and (\ref{Ey}).

\section{Modeling of metal-induced $p-n$ junctions in graphene}

Here, we present some simulations details of metal-induced $p-n$ junction in graphene, where the bulk part of graphene is controlled by the bottom global gate. We assume the source contact to be located at $x=0$, $z<0$, while the back gate is located at $x>0$, $z<-d$. Graphene occupies the half-plane $z=0$, $x>0$. The source contact dopes graphene, such that local electric potential at the source equals the built-in voltage, $V_{bi}$. The back gate is at potential $V_{bg}$. To simulate the doping, we present the full potential as external, $\phi_{ext}$ (created by contacts and gates), plus the potential created by induced electrons and holes in graphene, $\phi_{ind}$.
\begin{equation}
\label{Potential-definition}
    \phi(x) = \phi_{ext}(x)+\phi_{ind}(x).
\end{equation}
The external potential is found by applying a conformal mapping technique to the particular geometry of contacts, and is given by \cite{Exact}:
\begin{equation}
    \phi_{ext}(x) = \frac{2 V_{bi} }{\pi } \arctan\left(\frac{d}{x}\right)+ \frac{2 V_{bg}}{\pi } \arctan\left(\frac{x}{d}\right).
\end{equation}
The induced potential depends on the sheet density of electrons in graphene, $\rho_{2d}(x)$, and is given by
\begin{equation}
    \phi_{ind}(x) = \int_0^{\infty}{\rho(x')G(x,x')dx'},
\end{equation}
where $G(x,x')$ is the Green's function for the Poisson's equation. It can be found with the image-charge method, and is given by:
\begin{equation}
    G(x,x') = \frac{1}{\varepsilon_b}\ln \left[\frac{(x+x')^2 \left((2 d)^2+(x-x')^2\right)}{(x-x')^2 \left((2d)^2+(x+x')^2\right)}\right],
\end{equation}
where $\varepsilon_b = 4$ is the background dielectric constant (silicon dioxide) and $d$ is the gate-channel separation.

\begin{figure}[ht!]
    \includegraphics[width=0.9\linewidth]{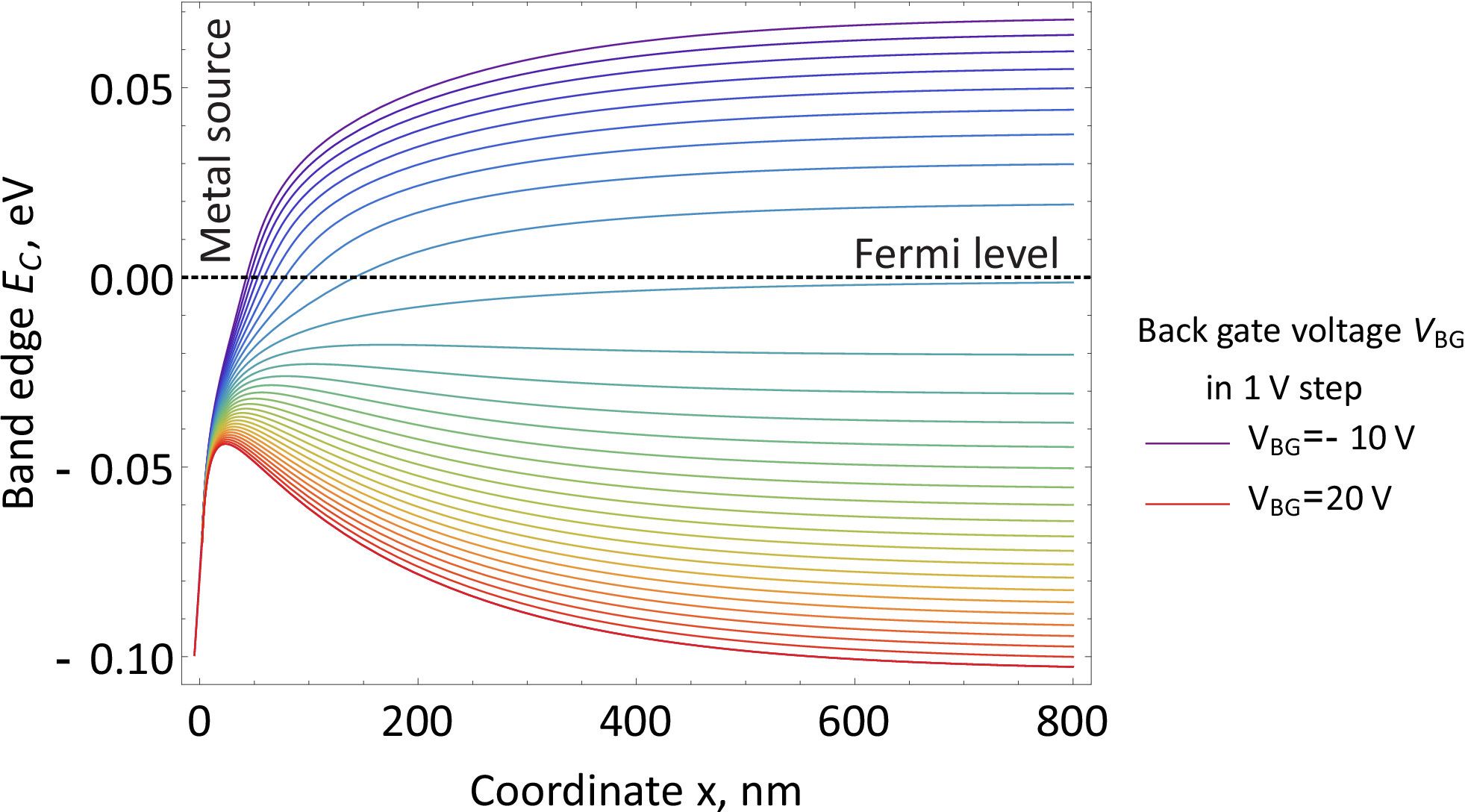}
    \caption{\label{figS1} Calculated coordinate dependence of band edge, $E_C(x)$, in graphene contacting with 'doping metal' at $x=0$. The back gate located at $d=500$ nm covers the entire graphene layer. Built-in voltage $V_{bi}=0.1$ V, temperature $T=300$ K, background dielectric constant $\varepsilon_b=4$.}
\end{figure}

Finally, to close the system of equations, we use the constant value of Fermi level $F$ at equilibrium. Setting $F=0$, we find that local Fermi energy in graphene is just the negative of local electric potential, $E_F(x) = -|e|\phi(x)$. On the other hand, the local charge density is related to the Fermi energy via Fermi statistics:
\begin{gather}
\label{Rho-microsopic}
    \rho(x) = |e|\left[p(E_F(x)) - n(E_F(x))\right],\\
    n(E_F) = \frac{1}{\pi \hbar^2 v_0^2} \int\limits_{0}^{\infty}{\frac{EdE}{1+\exp\left\{(E-E_F)/kT\right\}}},\qquad p(E_F) = n(-E_F).
\end{gather}
The final non-linear integral equation is formed by combining Eqs.~(\ref{Potential-definition})--(\ref{Rho-microsopic}). It was solved by a relaxation method, where the initial distribution of density was set equal to uniform gate-induced doping of graphene. The results of modeling are presented in Fig.~\ref{figS1} for $d=500$ nm, $V_{bi} = 0.1$ V (which corresponds to metal-induced electron doping), $T=300$ K. We see that in a wide range of gate voltages, corresponding to $p$-doping of graphene bulk, the $p-n$ junction is located at $l_{p-n}\sim 100$ nm. This justifies the estimate made in the main text.

\bibliography{references}